\documentclass{article} 
\usepackage{iclr2020_conference,times}

\usepackage{amsmath,amsfonts,bm}

\def\eqref#1{equation~\ref{#1}}

\def\1{\bm{1}}

\DeclareMathAlphabet{\mathsfit}{\encodingdefault}{\sfdefault}{m}{sl}
\SetMathAlphabet{\mathsfit}{bold}{\encodingdefault}{\sfdefault}{bx}{n}

\usepackage{hyperref}
\usepackage{url}
\usepackage{natbib}
\usepackage{graphicx}
\usepackage{authblk}

\title{Image Quality Transfer Enhances the\\ Contrast and Resolution of Low-Field MRI\\ in African Paediatric Epilepsy Patients}

\author[1,*]{Matteo Figini} 
\author[1,*]{Hongxiang Lin} 
\author[2]{Godwin Ogbole}
\author[3]{Felice D'Arco}
\author[1]{Stefano B. Blumberg}
\author[4,5]{David W. Carmichael}
\author[1,6]{Ryutaro Tanno}
\author[1,4]{Enrico Kaden}
\author[7]{Biobele J. Brown}
\author[7]{Ikeoluwa Lagunju}
\author[3,4]{Helen J. Cross}
\author[1,7]{Delmiro Fernandez-Reyes}
\author[1]{Daniel C. Alexander}

\affil[1]{Centre for Medical Image Computing and Department of Computer Science, University College London, UK}
\affil[2]{Department of Radiology, College of Medicine, University of Ibadan, Nigeria}
\affil[3]{Great Ormond Street Hospital for Children, London, UK}
\affil[4]{UCL Great Ormond Street Institute of Child Health, UK}
\affil[5]{Department of Biomedical Engineering, King's College London, UK}
\affil[6]{Machine Intelligence and Perception Group, Microsoft Research Cambridge, UK}
\affil[7]{Department of Paediatrics, College of Medicine, University of Ibadan, Nigeria}
\affil[*]{These two authors contributed equally}

\iclrfinalcopy 
\begin{document}

\begin{center}
\maketitle
\end{center}

\begin{abstract}
1.5T or 3T scanners are the current standard for clinical MRI, but low-field (\textless 1T) scanners are still common in many lower- and middle-income countries for reasons of cost and robustness to power failures. Compared to modern high-field scanners, low-field scanners provide images with lower signal-to-noise ratio at equivalent resolution, leaving practitioners to compensate by using large slice thickness and incomplete spatial coverage. Furthermore, the contrast between different types of brain tissue may be substantially reduced even at equal signal-to-noise ratio, which limits diagnostic value. Recently the paradigm of Image Quality Transfer has been applied to enhance 0.36T structural images aiming to approximate the resolution, spatial coverage, and contrast of typical 1.5T or 3T images.  A variant of the neural network U-Net was trained using low-field images simulated from the publicly available 3T Human Connectome Project dataset.\\
Here we present qualitative results from real and simulated clinical low-field brain images showing the potential value of IQT to enhance the clinical utility of readily accessible low-field MRIs in the management of epilepsy.
\end{abstract}

\section{Introduction}

Most hospitals in developed countries are equipped with at least one 1.5T MRI scanner, and 3T scanners have become increasingly common. In contrast, permanent magnet MRI scanners, which are limited to magnetic fields lower than about 1T, are still widely used in lower- and middle-income countries (LMICs), due to limited funds and lack of infrastructure including frequent power outages. Low-field (LF) MRI suffers from lower signal-to-noise ratio (SNR) compared to high-field (HF) at equivalent spatial resolution \citep{Marques2019}. To counteract the SNR reduction, practitioners commonly acquire images with non-adjacent thick slices to reduce the acquisition time and crosstalk artifacts. Moreover, the contrast between grey matter (GM) and white matter (WM) is usually worse than at high field (HF) even at equivalent SNR and spatial resolution. \par
Presurgical imaging in paediatric epilepsy is an example application where high-quality MRI can make a substantial clinical difference. In LMICs in the sub-Saharan region, clinical management of epilepsy is largely based on EEG and clinical assessment \citep{Lagunju2015}. Neurologists and neurosurgeons lack diagnostic quality imaging to establish effective clinical management leaving affected children with long-term refractory epilepsy. Indeed, high-quality MRI is critical for identification and management of refractory (pharmaco-resistant) epilepsy cases. Surgery is commonly applied in these cases, and the removal of the epileptogenic areas is effective in eliminating or reducing disability \citep{Ryvlin2014}. However, effective surgery relies on the localisation of those areas, which often correspond to very subtle abnormalities on MRI that are visible only at higher fields (e.g. focal cortical dysplasias, polymicrogyria \citep{Salmenpera2005,Sidhu2018}).The contrast between GM and WM is also inherently reduced in children, especially younger than 2 years old, which makes MRI detection of paediatric epilepsy lesions at LF even more difficult \citep{Tan2017}. \par
One way to enhance images when the acquisition resources are limited is to apply Image Quality Transfer (IQT), a machine-learning framework aiming at improving the quality of medical images by transferring information from high-quality references \citep{Alexander2017,Tanno2017,Blumberg2018}. This paradigm was recently adapted to improve the resolution and contrast of LF MRIs, using pairs of real HF and corresponding simulated LF images \citep{Lin2019}. The inverse mapping is then learned enabling one to estimate, from a LF MRI acquired in a LMIC clinic, the image that would have been obtained by a state-of-the-art HF scanner. Here we show its application in paediatric patients with epilepsy attending our neurology clinics in a LMIC country in the sub-Saharan region, with the aim of improving the spatial resolution and contrast of LF MRI, and hence their diagnostic value.

\section{Methods}
We use a recently proposed variant of U-Net \citep{Lin2019} to improve the contrast and spatial resolution of LF MRI.

\subsection{MRI Data}

High-resolution axial T1-weighted (T1w) images of 30 subjects were obtained from the publicly available Human Connectome Project (HCP) dataset \citep{Sotiropoulos2013}, acquired on a 3T Siemens Connectom scanner with an isotropic voxel size of 0.7 x 0.7 x 0.7 mm\textsuperscript{3}.
To evaluate image features at LF and to test the algorithm, we used T1w images acquired on healthy subjects and paediatric epilepsy patients using a 0.36T MagSense 360 MRI System scanner with a voxel size of 0.9 x 0.9 x 7.2 mm\textsuperscript{3}, with a slice thickness (ST) of 6 mm and 1.2 mm gaps. For further evaluations (see sections 2.2 and 2.4) we used T1w images acquired on a 1.5T Siemens Avanto with an isotropic voxel size of 1 x 1 x 1 mm\textsuperscript{3}.

\subsection{Simulation of Low-Field Images}

\begin{figure}[ht]
    \centering
    \includegraphics[width=1\textwidth]{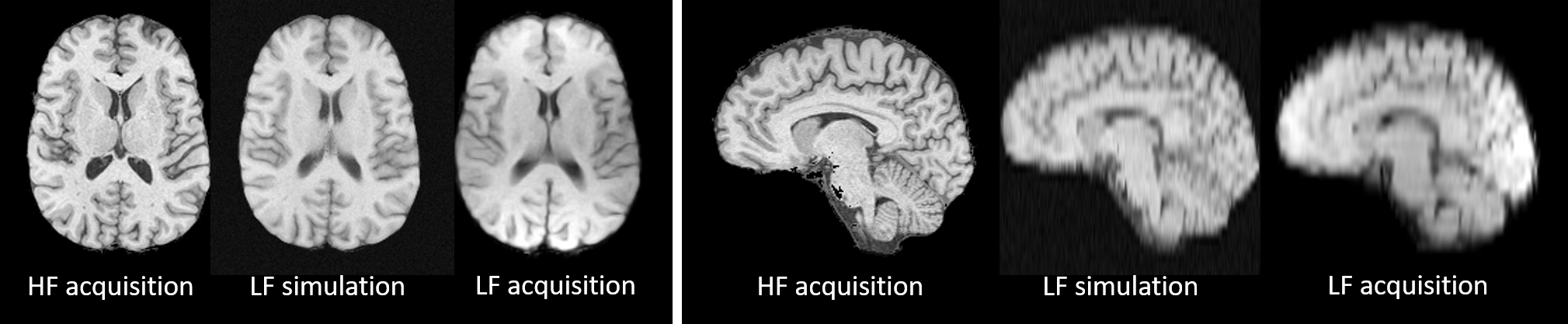}
    \caption{HF acquisition, corresponding LF simulation and LF acquisition for one of the healthy subjects with paired 0.36T and 1.5T scans. The simulations were performed with SNR values 48 for GM and 57 for WM and k = 6. With respect to the input HF acquisition, the output of the LF simulation resembles the real LF acquisition, with a loss of contrast and details in the axial (left) and especially in the sagittal view (right).}
    \label{fig1}
\end{figure}

Due to the difficulty in obtaining paired LF and HF MRIs from the same subjects, we simulated LF images from HF references, which has the advantage of avoiding misalignment issues that often affect patch-based image-to-image deep-learning \citep{Blumberg2019}. We defined a LF simulation procedure taking as inputs a HF image and the desired mean SNR for GM and WM in the output image. The simulation procedure includes the following steps:
\begin {enumerate}
    \item	Segmentation into GM, WM and CSF using the Unified Segmentation algorithm in SPM \citep{Ashburner2005}; the algorithm returns membership maps, with values between 0 and 1 defining the probability that each voxel belongs to each tissue (GM, WM or CSF).
    \item	Skull-stripping, i.e. exclusion of all the voxels with 0 probability of belonging to GM, WM or CSF.
    \item	Down-sampling: a Gaussian filter is applied along the slice direction, with a full-width at half maximum equal to the desired ST; values are then sampled every (ST+gap) mm. Note that brain sections inside the simulated gaps have virtually no effect on the output images.
    \item	Contrast change: target LF SNRs are estimated from example LF images; GM and WM multipliers are derived to get the appropriate ratio between the mean signal in GM and WM. Then the HF signal in each voxel is multiplied by a weighted average of the GM and WM multipliers, using the GM and WM membership values from the segmentation in step 1 as weights.
    \item Noise addition: Gaussian noise is added with an appropriate variance to get the desired mean SNRs.
\end{enumerate}
We validated the simulations using 3 sets of paired 0.36T and 1.5T images acquired on the same subjects; simulations were performed on the HF images using the SNR values measured at LF and the outputs were visually compared to the real LF images (Figure \ref{fig1}).

To create the training set, we measured the mean SNR in GM and WM in a representative set of 28 LF images and used these values for the simulations from HCP data (see next section 2.3).

\subsection{Deep Learning Framework and Training}

We used a variant of U-Net called Aniso-U-Net, proposed in \cite{Lin2019}. This architecture allows input and output to have the same in-plane spatial resolution but a different slice thickness; thus, it suits the purpose of anisotropic super-resolution, with an up-sampling factor k in the slice direction.
Paired HF and simulated LF images from 12 HCP subjects were used for training. For the simulations, the mean GM and WM SNR values from the LF dataset were used and the slice thickness and gap were 6 and 2 times the in-plane resolution, respectively, giving an overall down-sampling factor of k = 8. The loss function was the mean voxel-wise square error between the output and the HF ground truth.
Validation was performed on 3 additional HCP subjects in each epoch of the training phase, and 15 more were used for evaluation. Cubic B-spline interpolation was also performed in the evaluation images for comparison.

\subsection{Evaluation on patients’ images}

We evaluated the trained model in two clinical cases, both unseen during the training phase. The first one is an 18-year-old epilepsy patient with hippocampal sclerosis. Here HF T1w images showed a volume reduction of the left hippocampus; we simulated LF images, making volumetric evaluations virtually impossible, and tried to recover the information in the slice direction. The second case is a 10-year-old epilepsy patient who had two cystic lesions at the GM-WM junction of the parietal lobes, that were clearly visible on T2-weighted (T2w) images but only slightly hypointense on T1w images at LF; here we aimed to enhance the lesion conspicuity on T1w images using the T2w as a reference.
The images were inspected by two neuroradiologists (1 paediatric neuroradiologist with 6 years of clinical experience on epilepsy images and 1 adult and paediatric neuroradiologist with 10 years’ experience in standard and LF neuroimaging). In particular, the GM-WM interface was evaluated in the original and enhanced scans.

\section{Results}

\subsection{Evaluation on HCP subjects}

In the HCP evaluation set, the average structural similarity index for the IQT-enhanced images was 0.852 and significantly higher than for the interpolated images (0.567).

\subsection{Evaluation in epilepsy patients – simulated LF images}

\begin{figure}[ht]
    \centering
    \includegraphics[width=0.7\textwidth]{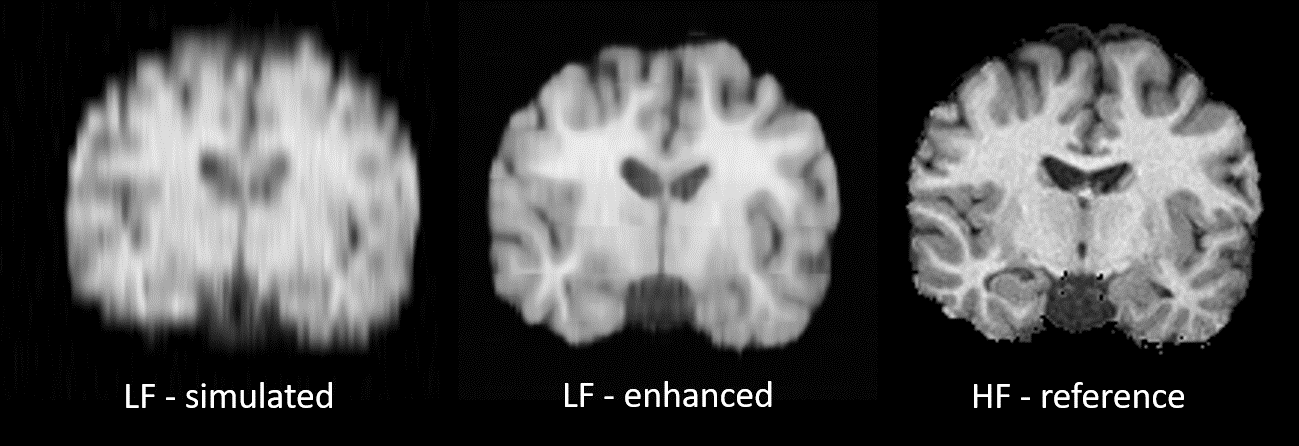}
    \caption{Coronal views of the IQT results for an 18-year-old patient with hippocampal sclerosis. The LF image on the left was simulated from the real HF acquisition on the right with SNR values 50 for GM and 63 for WM and k = 8, leading to a significant loss of detail outside the axial plane. The IQT-enhanced image in the centre showed a remarkable visual improvement.}
    \label{fig2}
\end{figure}

In the first clinical case, the GM-WM boundaries were barely visible in the coronal view (commonly used by radiologists for the hippocampus) of the simulated LF image and it was difficult to even locate the hippocampus. IQT allowed a remarkable visual improvement of the coronal image, showing GM-WM boundaries much more clearly and allowing an approximate evaluation of hippocampal volume (Figure \ref{fig2}).

\subsection{Evaluation in epilepsy patients – real LF images}

\begin{figure}[ht]
    \centering
    \includegraphics[width=1\textwidth]{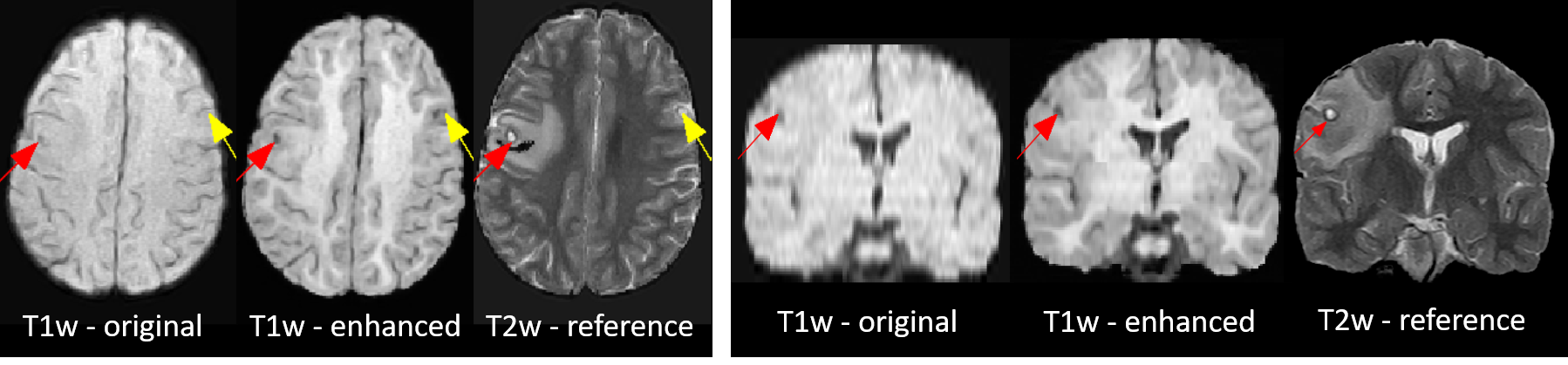}
    \caption{Original and IQT-enhanced images from a 10-year-old epilepsy patient with cystic lesions, shown in axial (left) and coronal view (right) \citep{Lin2019}. The larger cyst (red arrow) is surrounded by extensive edema and its conspicuity on T1w images is enhanced by IQT. The smaller cyst (yellow arrow) is almost invisible on the original T1w and becomes evident after IQT. Axial and coronal LF T2w images are shown as reference; note that they come from two separate acquisitions, whereas the T1w was only acquired axially.}
    \label{fig3}
\end{figure}

In the second case the appearance of the two cystic lesions was greatly enhanced by IQT, particularly the left one (yellow arrow in Figure \ref{fig3}) was not visible in the original T1w examination and clear in the enhanced T1w image. The GM-WM interface was judged, in consensus, to be definitely sharper in IQT enhanced images (Figure \ref{fig3}).

\section{Discussion and Conclusion}

We showed preliminary results in the enhancement of LF MRI images by IQT in paediatric epilepsy patients. Our IQT algorithm is based on Aniso-U-Net trained on synthetic LF images simulated from real HF data by imposing the SNR and spatial resolution of a representative set of real LF scans. The results show an improvement of both resolution and contrast of clinical LF images, allowing a better radiological evaluation of subtle lesions and abnormalities in paediatric patients with epilepsy. \par
In one case we had HF images and simulated the LF ones. This allowed us to test the algorithm in an idealized scenario where the test data is generated in the same way as the training data and a paired HF ground truth is available.  We were able to improve the image contrast and detail, in particular at the GM-WM boundaries, in the orientations perpendicular to the simulated acquisition plane.\\
In a second case we used real LF scans and we were able to enhance the visibility of two cystic lesions on T1w images. We had no HF reference, but the lesions were clearly visible on T2w images. This finding is relevant since T1w rather than T2w images are more useful for identification of focal cortical dysplasias in a relevant percentage of patients with intractable seizures. Also, diagnostic performance is generally enhanced when contrast changes are visible in images with different contrast; this can increase confidence in their identification and better define the likely pathological substrate. \par
Our simulation approach is to directly impose the SNR and contrast estimated in real LF images; more sophisticated methods exist in literature, but they did not prove effective or computationally feasible for this application. Biophysical models \citep{Wu2016} depend on the knowledge of the magnetic field dependence of relaxation times (T1, T2 and T2*), which is tissue-dependent and has been investigated only in a limited number of studies and in a relatively narrow range of magnetic fields \citep{Marques2019}. Moreover, some features of real acquisition setups (e.g. coil sensitivity, field homogeneity, pre-processing and filtering) may be difficult to fully model.\\
On the other hand, learning-based methods (e.g. \cite{Freeman2000}) have heavy computational cost and require large paired datasets, which may not be justified for this sub-step of IQT. They should be rather considered as an alternative to the whole IQT process, if a large enough paired dataset is available for training. \\
We purposely kept our algorithm simple for reasons of robustness, especially important when dealing with pathological image alterations. We focused on the image features that are most relevant for the radiological evaluation of subtle lesions in epilepsy patients, namely the contrast between GM and WM and the SNR, as well as spatial resolution. Further work will be devoted to including in the simulations more subtle changes between HF and LF and imaging artifacts. \par
Current work is on-going to evaluate the proposed algorithm in a more extensive set of clinical images with multiple contrasts, and to assess the added value of the enhanced images from a perspective of clinical decision support. We will also need to understand to what extent pathological alterations should be included in the training dataset to get generalizability but prevent the model hallucinating non-existing abnormalities \citep{Cohen2018}. \\
IQT might also be combined with advanced acquisition and reconstruction methods, such as compressed sensing \citep{Lustig2007}, though these would require access to the acquisition sequences and a technical expertise that may not be widely available in LMICs. \par
IQT, apart from contributing significantly in enabling improved clinical decision making for children with epilepsy in Africa, holds potential to transform the imaging landscape in other clinical conditions where diagnostic accuracy and improved image quality are essential in LMICs with limited imaging infrastructure and no access to cutting-edge technology.

\subsubsection*{Acknowledgments}
This work was supported by EPSRC grants (EP/R014019/1, EP/R006032/1 and EP/M020533/1), the NIHR UCLH Biomedical Research Centre and the NIHR GOSH Biomedical Research Centre. Data were provided in part by the Human Connectome Project, WU-Minn Consortium (Principal Investigators: David Van Essen and Kamil Ugurbil; 1U54MH091657) funded by NIH and Washington University. The 0.36T MRI data were acquired at the University College Hospital, Ibadan, Nigeria. The clinical 1.5T data used for the evaluation of simulated images (section 3.2) were acquired at the Great Ormond Street Hospital for Children, London, UK.

\bibliography{IQT_Nigeria_ICLR2020_AI4AH}
\bibliographystyle{iclr2020_conference}

\end{document}